\pdfoutput=1 
\documentclass[conference]{IEEEtran}
\IEEEoverridecommandlockouts
\usepackage{cite}
\usepackage{amsmath,amssymb,amsfonts}
\usepackage{algorithmic}
\usepackage{graphicx}
\usepackage{textcomp}
\usepackage{color,soul}
\usepackage{caption}
\usepackage{subcaption}
\usepackage{booktabs} 
\usepackage{makecell} 
\usepackage[flushleft]{threeparttable}     
\usepackage{xcolor}
\def\BibTeX{{\rm B\kern-.05em{\sc i\kern-.025em b}\kern-.08em
    T\kern-.1667em\lower.7ex\hbox{E}\kern-.125emX}}
\begin{document}

\title{Human-Machine Interface Evaluation Using EEG in Driving Simulator}

\author{\IEEEauthorblockN{1\textsuperscript{st} Yuan-Cheng Liu}
\IEEEauthorblockA{\textit{Chair of Ergonomics} \\
\textit{Technical University of Munich}\\
Munich, Germany \\
yuancheng.liu@tum.de}
\and
\IEEEauthorblockN{2\textsuperscript{nd} Nikol Figalova}
\IEEEauthorblockA{\textit{Department of Clinical and health Psychology} \\
\textit{Ulm University}\\
Ulm, Germany \\
nikol.figalova@uni-ulm.de}
\and
\IEEEauthorblockN{3\textsuperscript{rd} Martin Baumann}
\IEEEauthorblockA{\textit{Department of Human Factors} \\
\textit{Ulm University}\\
Ulm, Germany \\
martin.baumann@uni-ulm.de}
\and
\IEEEauthorblockN{4\textsuperscript{th} Klaus Bengler}
\IEEEauthorblockA{\textit{Chair of Ergonomics} \\
\textit{Technical University of Munich}\\
Munich, Germany \\
bengler@tum.de}

}

\maketitle

\begin{abstract}

Automated vehicles are pictured as the future of transportation, and facilitating safer driving is only one of the many benefits. However, due to the constantly changing role of the human driver, users are easily confused and have little knowledge about their responsibilities. Being the bridge between automation and human, the human-machine interface (HMI) is of great importance to driving safety. This study was conducted in a static driving simulator. Three HMI designs were developed, among which significant differences in mental workload using NASA-TLX and the subjective transparency test were found. An electroencephalogram was applied throughout the study to determine if differences in the mental workload could also be found using EEG's spectral power analysis. Results suggested that more studies are required to determine the effectiveness of the spectral power of EEG on mental workload, but the three interface designs developed in this study could serve as a solid basis for future research to evaluate the effectiveness of psychophysiological measures.

\end{abstract}


\section{Introduction}\label{intro}

Automated vehicles (AV) have been considered a revolutionary technology for being economical and environmentally friendly, efficient in transportation, and able to increase driving safety \cite{Smith2015}. To facilitate this, the interaction between users and AV has been found pivotal \cite{Koeber2013}, making the human-machine interface (HMI) on AV crucial for users to operate it properly. Various studies regarding in-vehicle or external HMI designs have been conducted \cite{Richardson2018, Rettenmaier2019, Bengler2020},

However, some researchers argued that current HMI designs are prone to error and confusing for users \cite{Carsten2019}. Whenever the transitions between different automation levels are made, the distribution of responsibilities between users and AV constantly changes. This could lead to confusion for users and even cause accidents while driving. To solve the problems, it would be essential for HMI designs to be transparent, i.e., transmit the information correctly, efficiently, and understandably.

There are currently numerous guidelines and evaluation methods aiming to help develop a transparent HMI. Still, most studies are either heavily based on experts' perspectives or gathered heuristically and subjectively. Some researchers developed a guideline to design and verify AV HMIs with a thorough itemized checklist to help evaluate whether the HMI design has fulfilled the recommendations \cite{Naujoks2019}. Similarly, some researchers analyzed the HMI design based on usability heuristics and came up with suggestions to improve the HMI design \cite{Parkhurst2019}. These methods could provide valuable aspects and help improve HMI designs, but using only subjective and heuristic evaluations would not be enough. Hence, a standardized, objective, and efficient way for HMI design evaluation is urgently required to approach the optimal design for all sorts of different scenarios, levels of automation, and user characteristics.

In the previous study, we developed a standardized transparency assessment test to evaluate the HMI designs that were already available on the market. We evaluated how easy the HMI designs were for users to understand the information transmitted correctly \cite{Liu2022}. The resulting TRASS (the proposed Transparency Assessment Method) was calculated by the user's answer accuracy of the given HMI state, and the estimated workload score based on the time to understanding (i.e., the time took to start answering questions). The results show that the proposed method effectively found the difference among the HMI designs, and was validated in an online study. However, the applicability is limited to static scenarios and needs exploration in a higher-fidelity environment. To do so, we need an alternative way to measure the workload in real time and still do it objectively. By doing so, the HMI design process could be handled in a more systematic way, and in return, be more efficient.

\begin{figure}[htbp]
     \centering
     \includegraphics[width=0.45\textwidth]{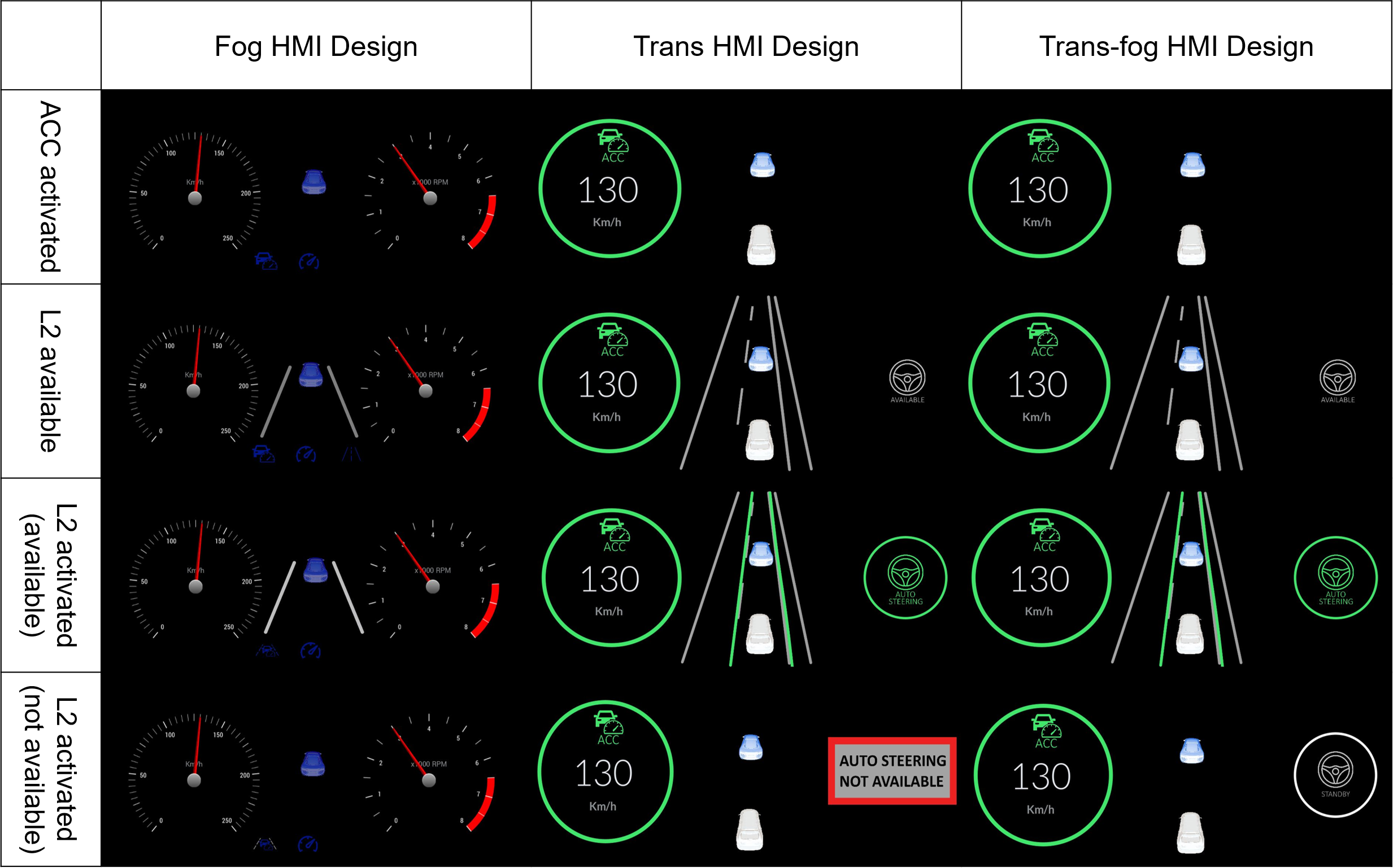}
     \caption{HMI designs under given circumstances.}
     \label{fig:table_HMI_designs}
\end{figure}

In this paper, we intend to evaluate the effectiveness of the electroencephalographic (EEG) in identifying differences in mental workload while interacting with different HMI designs during simulated driving. This evaluation would also be the first step in extending the proposed assessment method into a real driving scenario and developing a real-time HMI assessment method. We first developed three different SAE Level 2 (L2) HMI designs \cite{SAE2021}, as in Fig.~\ref{fig:table_HMI_designs}, which had different transparencies according to the results from the previous study \cite{Liu2022}, and should result in different mental workloads during the interaction. To validate this idea, differences in workload among three HMI designs were first evaluated using subjective workload measures. Then, evaluations using the power spectral analysis of the EEG data were conducted, and comparisons between the EEG results and subjective workload measurements were also made.

The research question is, whether any differences in mental workload among the HMI designs developed exist and if they could be found using analysis of the spectral power of EEG. Thus the corresponding hypotheses are set as follows:

\begin{itemize}

    \item H1: Differences in mental workloads could be found among different HMI designs using NASA-TLX.
    \item H2: Differences in mental workloads could be found among different HMI designs using a subjective transparency questionnaire.
    \item H3: Differences in mental workloads could be found among different HMI designs using the spectral power of EEG.
\end{itemize}

\section{Method}\label{method}

\subsection{Human-machine interface designs}
To evaluate whether EEG data is capable of telling the difference in mental workload among different HMI designs, we first need HMI designs that are distinct in understandability, which would result in different mental workloads. Owing to the fact that there hasn't been a standardized evaluation system to determine how understandable the HMI is, a heuristic approach is still needed at this stage. We followed the design principles for human-computer interface and also results from the previous study \cite{Norman1983, Liu2022}, we developed the following three in-vehicle HMI that should be distinct in HMI understandability and correctness on information transmitted, i.e., HMI transparency:

\subsubsection{Fog HMI design}
The fog HMI design, as in the left column of Fig.~\ref{fig:table_HMI_designs}, represents the HMI design that contains all the design elements which would mitigate the clarity of the information transmitted and hence diminish the HMI transparency for users. Looking at the HMI transparency side, those disadvantageous elements include small icons, which indicate levels of automation and whether the sensor is functioning properly, low contrast color, and redundant icons for the same function. These elements either violate the design principle or are pointed out by the participant in the last study stating that they were confusing and misleading. On the side of system transparency, there would be no indication of failing to activate L2 automation, and it would go straight into standby mode. That is if the L2 automation is activated by the user. Still, the activation failed due to the system limit being reached, there would be no indication that the activation failed and that the AV is not controlling the vehicle laterally.

\subsubsection{Trans HMI design}
The Trans HMI design, as in the middle column of Fig.~\ref{fig:table_HMI_designs}, represents the HMI design that contains all the design elements which would intensify both the HMI and system transparencies. On the HMI transparency side, it fulfills all the design principles, including high contrast color, large icons, and no redundant icons, in contrast to fog HMI design. Regarding system transparency, there is no standby mode, which means if the L2 automation cannot be activated, the system would show a warning on the HMI to remind the users that the L2 automation failed to activate and that they should still control the steering (as shown in the rectangle at the bottom of the middle column of Fig.~\ref{fig:table_HMI_designs} ).

\subsubsection{Trans-fog HMI design}
The trans-fog HMI design, as in the right column of Fig.~\ref{fig:table_HMI_designs}, represents the HMI design with the HMI transparency as Trans HMI design, but with minimum system transparency as fog HMI design. It satisfies all the design principles like Trans HMI design and has the same icons and colors. The only difference is when L2 automation is activated but turns out to be unavailable. In a scenario like this, the trans-fog HMI design would go into standby mode like the fog HMI design, resulting in minimum system transparency.


\subsection{Psychophysiological and self-reported measures}
In this study, multiple measures are used to evaluate users' cognitive state while interacting with the HMI designs, including psychophysiological and self-reported measures. When evaluating human cognitive states, using only the self-reported method has shown to be less accurate, as participants are usually not precise when it comes to judging their own cognitive state. Psychophysiological measures also have the advantage over other objective methods like behavioral measures. Since those physiological events are not under voluntary control, they could avoid being affected by unrelated constructs like drowsiness or stress during driving. It also has been suggested that psychophysiological measures could significantly increase the accuracy of mental workload measurements \cite{Yang2016}.

The EEG signal has been found to be a helpful psychophysiological measure when evaluating mental workload in driving scenarios \cite{Pollmann2019}, as it provides a non-intrusive way to measure neural activities while also allowing the high temporal resolution to acquire the data in real-time. In this study, we analyzed the oscillatory brain activity derived from EEG signals to determine its relation to specific neurocognitive functions. For instance, decreased alpha power activity (8-12 Hz) and increased theta power activity (4-7 Hz) are often associated with increased mental workload \cite{Mun2017}.

To the best of our knowledge, there has not been a study to investigate different workloads intrigued by HMI designs with different transparency, so we also included the self-reported measure (i.e., NASA-TLX) as a reference and compared it to results in the literature. 

\subsubsection{EEG signal recording and pre-processing}
The EEG was recorded using 32 channels electrodes placed according to the international 10-20 system. ActiCAP set (Brain Products GmbH, Germany) was used, with active shielded electrodes and a LiveAmp amplifier. The data were recorded with a 1000 Hz sampling frequency and preprocessed in Matlab version R2022a. 

We created an alternative dataset to perform adaptive mixture independent component analysis (AMICA). The alternative datasets were first downsampled to 500 Hz and bandpass-filtered between 0.1 Hz and 100 Hz. Line noise artifacts were removed using the ZapLine plugin \cite{DeChev2020, Klug2022}. Channels that correlated with their own robust estimate less than r = .78 more than 50 percent of the time were interpolated, and the data were re-referenced to the common average. We then applied AMICA as the blind source separation using ten iterations \cite{Palmer2008}. The spatial filter produced by AMICA was then copied into the original, raw dataset, and the components were calculated using IClabel. We used the popularity classifier, which removes components that most likely do not originate in brain activity. Finally, we used a second-order Butterworth filter and bandpass filtered the data between 0.5 Hz and 30 Hz.

The pre-processed data were then epoched into fragments from -10 s to 10 s with regards to the activation of L2 automation, which was the time slot during which participants had to focus on the HMI design in order to do the transparency test that followed. Afterwards, fast Fourier transformations were applied to obtain the spectral power distribution. The relative mean spectral power of the alpha band (8-12 Hz) from bi-lateral parietal electrodes (Pz, P3, P4) and the relative mean spectral power of the theta (4-7 Hz) from bi-lateral frontal electrodes (Fz, F3, F4) were calculated with respect to the total power (0.5-30 Hz) \cite{Pollmann2019}.

During the spectral power analysis, all the 20-second epochs gathered with the same HMI design were combined for each participant. This resulted in one 80-second EEG pre-processed data for each HMI design, which was later used for the final statistic analysis.

\subsubsection{Subjective evaluations}
The first subjective evaluation was the mental workload evaluated by the National Aeronautics and Space Administration-Task Load Index (NASA-TLX) \cite{nasatlx}, which is a six-item questionnaire. In this study, the NASA-TLX scores were calculated without weighted parameters, so we averaged scores across six items. Besides, since each HMI design was measured four times for each participant, we also need to calculate the average again. The final scores calculated were then used as the subjective workload of each HMI design of the participant. 

The second subjective evaluation consisted of three questions regarding HMI transparency and was derived from the TRASS test in the previous study \cite{Liu2022}. First, participants were asked to evaluate whether they agreed that they could understand the HMI design. Then, they were asked if they agreed that they could obtain critical information from the HMI design. Lastly, they were asked whether they agreed that the HMI design was easy to understand. All three questions were scaled from 0 to 100.

\subsection{Participants}
Twelve participants were recruited for this study, where three were male, eight were female, and one was diverse. The ages ranged from 22 to 30, with mean age $= 26.92$, $SD = 3.87$. The data from two of the participants were excluded due to corrupted recordings. All of the participants came with valid driving licenses, and they had held them for at least three years ($M = 8.83, SD = 3.71$)

\subsection{Procedures}
\begin{figure}[htbp]
     \centering
     \includegraphics[width=0.45\textwidth]{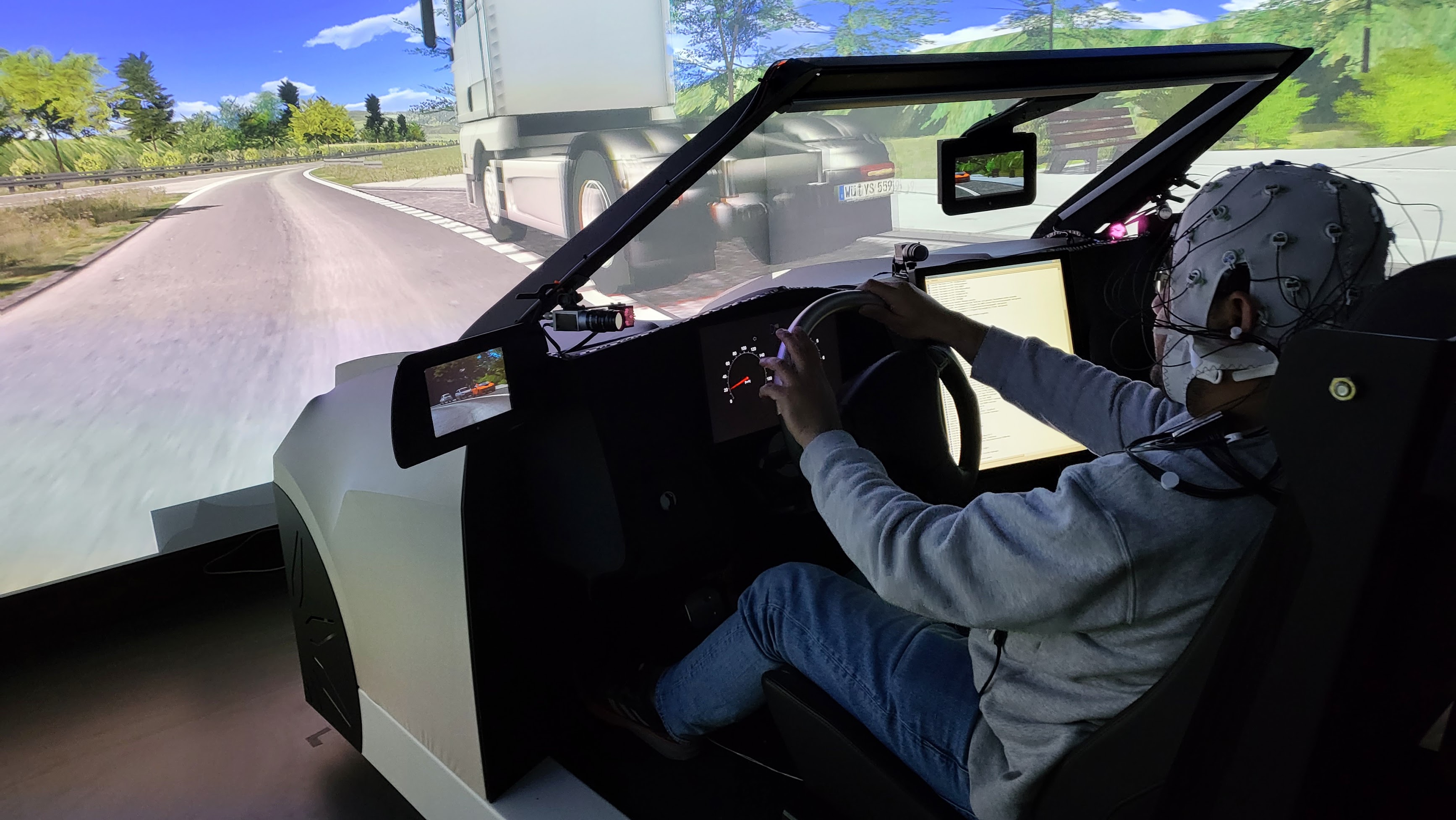}
     \caption{Illustration of participant with EEG set up in the driving simulator.}
     \label{fig:simulator}
\end{figure}

The study was conducted in a static driving simulator with a field of view of 120$^{\circ}$, as shown in Fig.~\ref{fig:simulator}. The front panel consists of three screens with the scenes projected from three projectors respectively. It also has rear, left, and right mirrors, which are small LCD displays. The software SILAB was used to create the 4-lanes highway scenario. A touch screen was fixed on the right-hand side of the driver's seat, where automated cruise control (ACC) and L2 automation activation buttons were shown.

Upon arrival, participants were welcomed and briefly introduced to the study. After finishing the demographic questionnaire and setting up the EEG electrodes, a pre-recorded video containing detailed instructions was played. Then, participants were brought to the driving simulator and started a familiarization test drive. During the test drive, the baseline EEG data were also collected. After the familiarization, the formal test would begin if no symptoms of simulator sickness were shown.

\begin{figure}[htbp]
     \centering
     \includegraphics[width=0.45\textwidth]{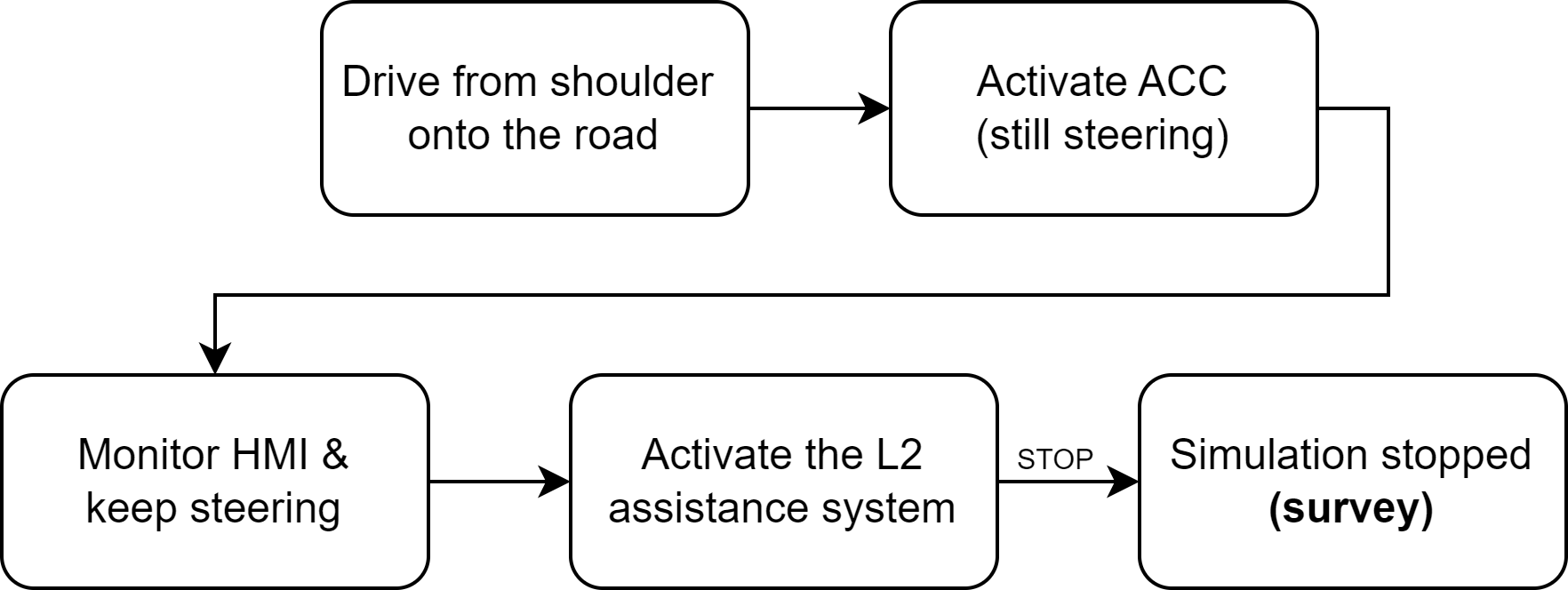}
     \caption{Experimental procedure during each trial.}
     \label{fig:procedure}
\end{figure}

Each participant had to follow the procedure in each trial, as shown in Fig.~\ref{fig:procedure}. Five seconds after the activation of the L2 automation (regardless of whether it was successfully activated or not), the simulation ended with a fade-out animation to avoid simulator sickness, and the participant was asked to complete a survey consisting of NASA-TLX and subjective transparency test. Participants had to go through all three HMI designs in counterbalanced order, where each HMI design had four different traffic layouts, making a total of 12 trials for each participant. The EEG signal was recorded throughout the study. Finally, participants were compensated at the end of the study after the feedback section about the procedure, questionnaires, and HMI designs.

\section{Results}\label{result}
To evaluate whether three different HMI designs had any effects on mental workload during driving, we conducted the repeated measure ANOVA to test the significance of the differences in the NASA-TLX and EEG data. The same approach was also used for subjective and objective transparency tests. Multiple comparisons were made with post hoc analysis (Holm's).

\subsection{Subjective evaluations}
\begin{figure}
     \centering
     \begin{subfigure}[htbp]{0.232\textwidth}
         \centering
         \includegraphics[width=\textwidth]{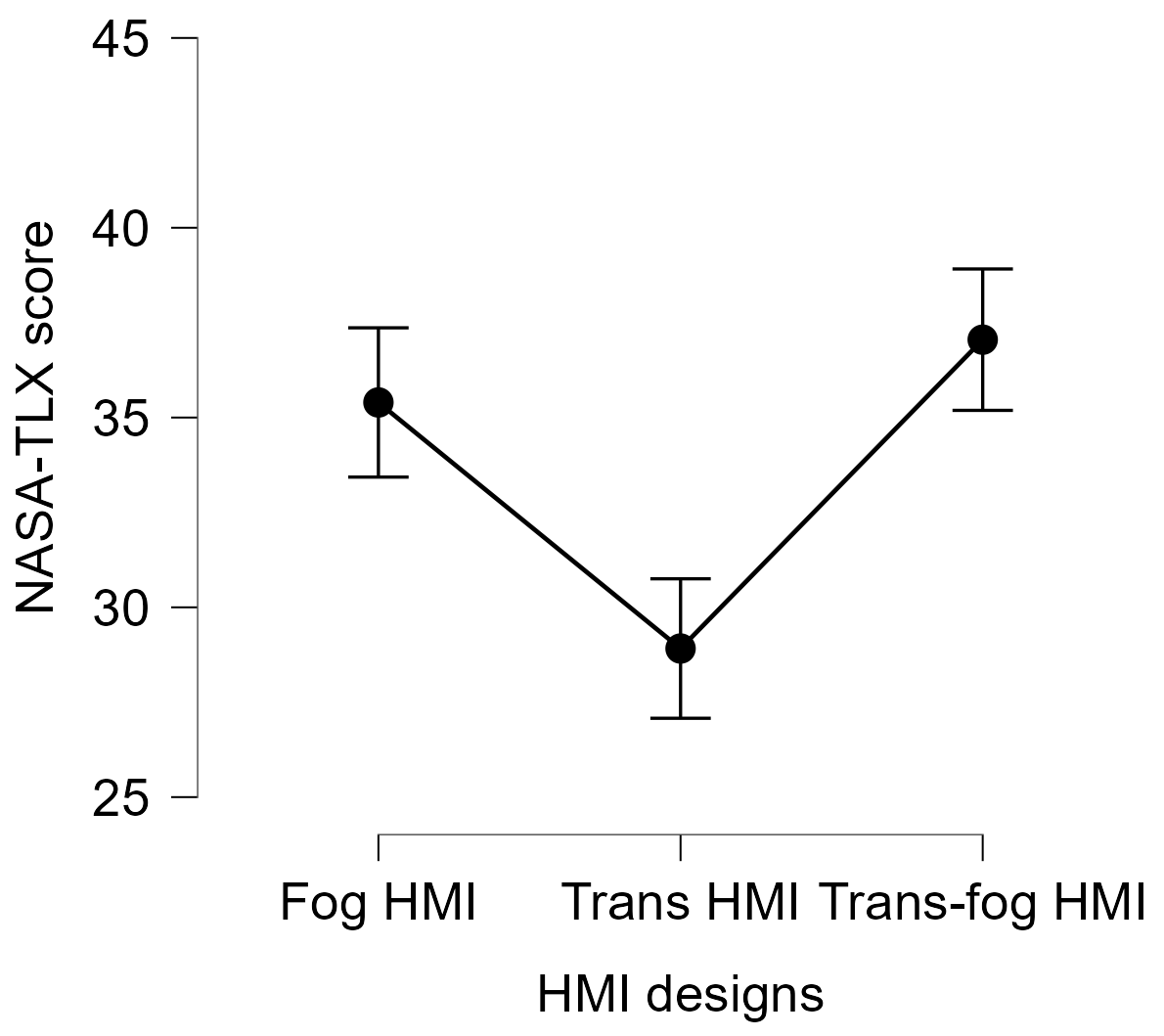}
         \caption{NASA-TLX score.}
         \label{fig:nasatlx}
     \end{subfigure}
     \hfill
     \begin{subfigure}[htbp]{0.25\textwidth}
         \centering
         \includegraphics[width=\textwidth]{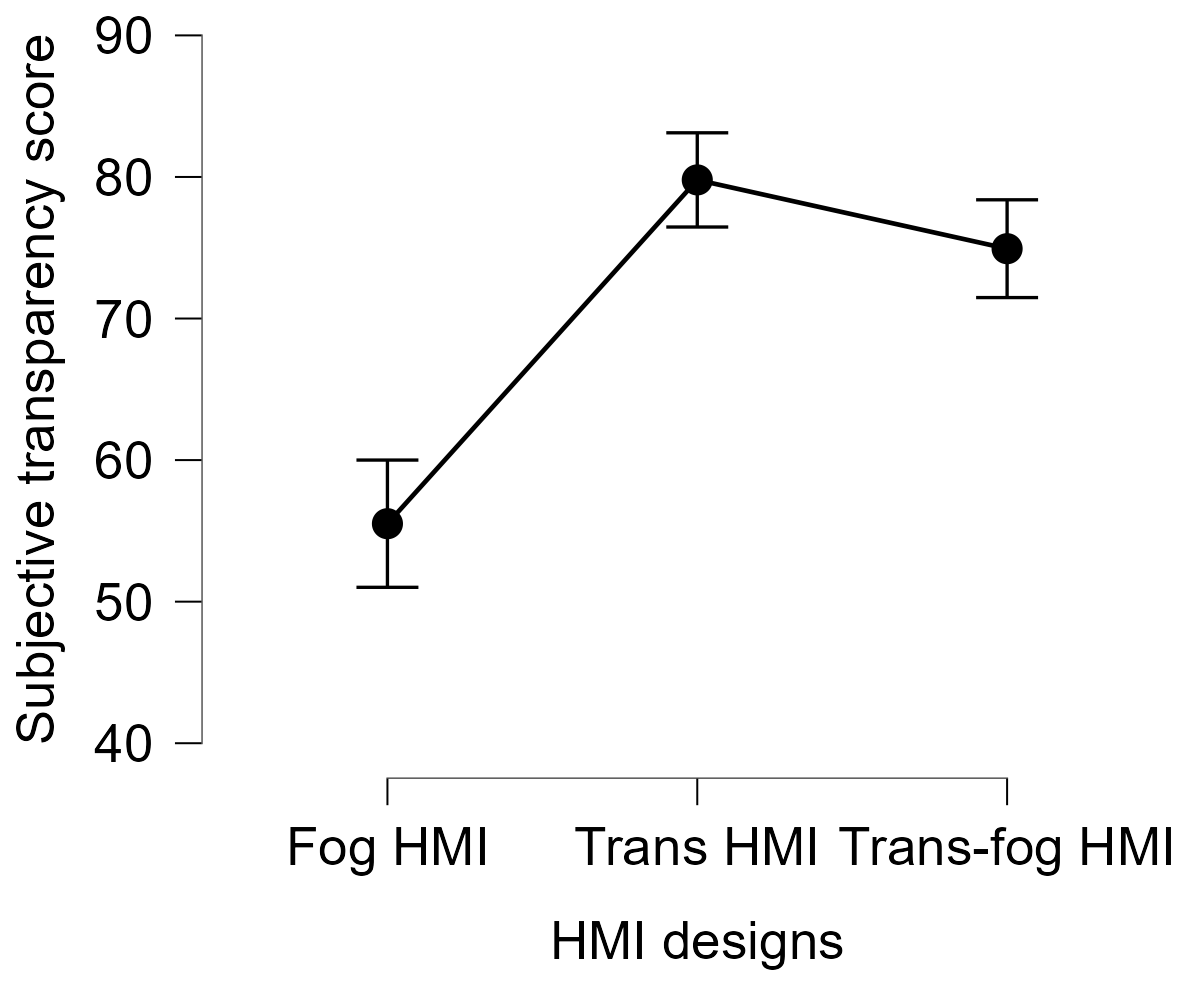}
         \caption{Subjective transparency score.}
         \label{fig:sub_trans}
     \end{subfigure}
        \caption{Subjective evaluations with mean values and standard errors of means.}
        \label{fig:subjective_result}
\end{figure}

Fig.~\ref{fig:subjective_result} shows the resulting NASA-TLX and subjective transparency scores. The averaged NASA-TLX scores were found significantly different among those three HMI designs $F(2,78)=5.18, p=0.008, \eta_{p}^{2}=0.12$, where Trans HMI had the lowest score among them. A similar outcome was found on subjective transparency scores, where the effect of HMI designs was found to be significant $F(2,78)=11.47, p<0.001, \eta_{p}^{2}=0.56$.

To gain a better understanding of the relationships among the three HMI designs, post hoc comparisons were conducted for both NASA-TLX and subjective transparency scores. We can see from Table.~\ref{table:posthoc} that, the Fog HMI possessed a significantly higher mental workload and significantly lower subjective transparency when comparing it to the Trans HMI. However, when compared to the Trans-fog HMI, the significance was only found in its lower subjective transparency. For the Trans and Trans-fog HMIs, no difference in subjective transparency was found, but in NASA-TLX scores, significantly lower mental workload for the Trans HMI was identified.

\begin{table}[htbp!]
\centering
\begin{threeparttable}
\caption{Post Hoc Comparisons on HMI Designs for Subjective Evaluations.}
\label{table:posthoc}
\begin{tabular*}{0.45\textwidth}{@{\extracolsep{\fill}\quad}llcccc}
\toprule
\multicolumn{2}{c}{} & \multicolumn{2}{c} {} & \multicolumn{2}{c}{\textbf{Subjective}} \\
\multicolumn{2}{c}{} & \multicolumn{2}{c} {\textbf{NASA-TLX}} & \multicolumn{2}{c}{\textbf{transparency}} \\
\cmidrule(rl){3-4} \cmidrule(rl){5-6}
  &   & Cohen's d & $p_{holm}$ & Cohen's d & $p_{holm}$ \\
\midrule
\makecell[c]{Fog} & Trans & 0.33 & \textbf{0.035} & -1.37 & \textbf{\textless{}0.001}  \\
\makecell[c]  & Trans-fog & -0.084 & 0.538 & -1.097 & \textbf{0.006}  \\
\makecell[c] {Trans} & Trans-fog & -0.42 & \textbf{0.01}  & 0.27 & 1.00 \\
\bottomrule
\end{tabular*}
\end{threeparttable}
\end{table}

\subsection{Mental workload measurement with EEG}
\begin{figure}
     \centering
     \begin{subfigure}[htbp]{0.239\textwidth}
         \centering
         \includegraphics[width=\textwidth]{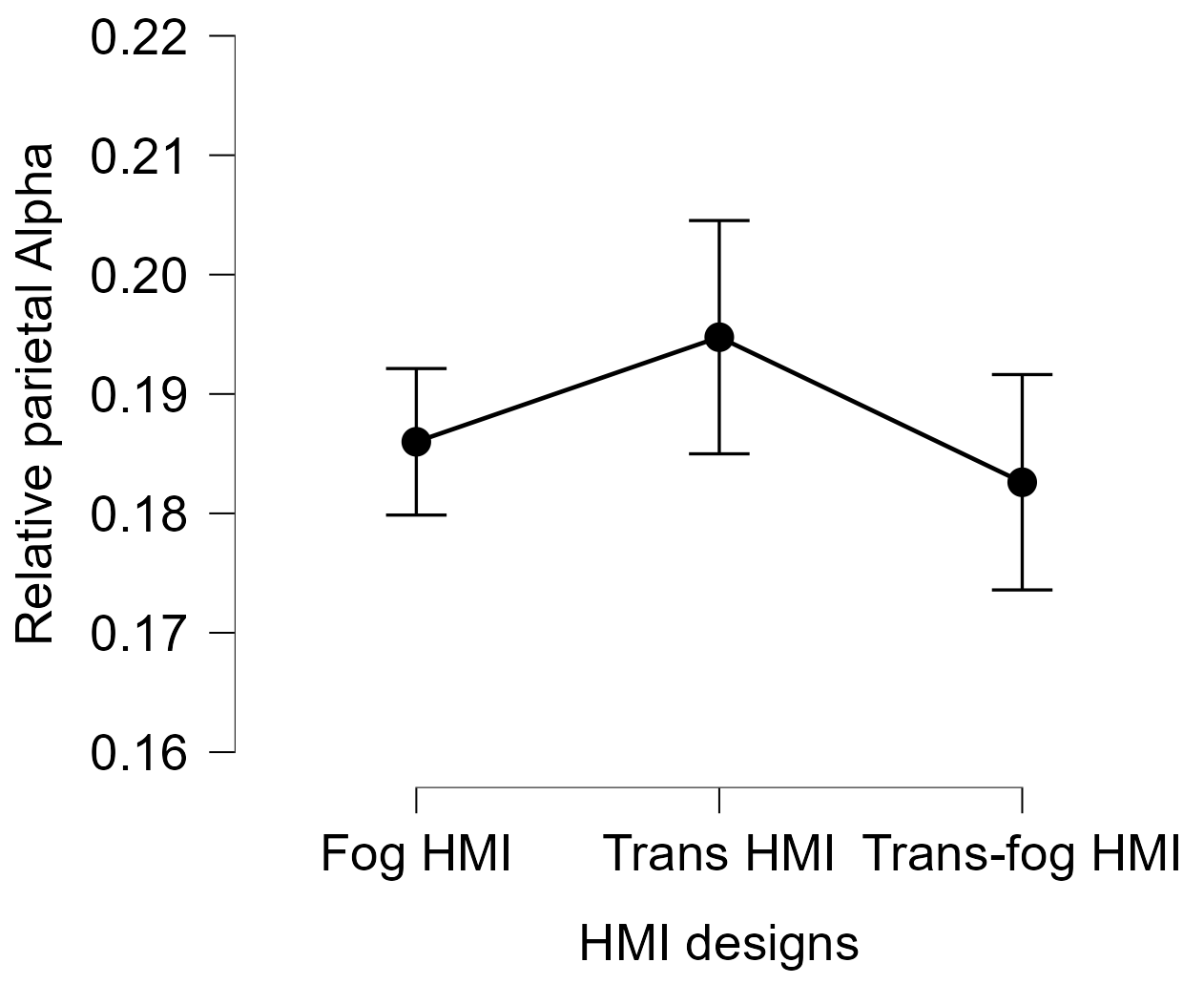}
         \caption{Relative Alpha power.}
         \label{fig:alpha_power}
     \end{subfigure}
     \hfill
     \begin{subfigure}[htbp]{0.242\textwidth}
         \centering
         \includegraphics[width=\textwidth]{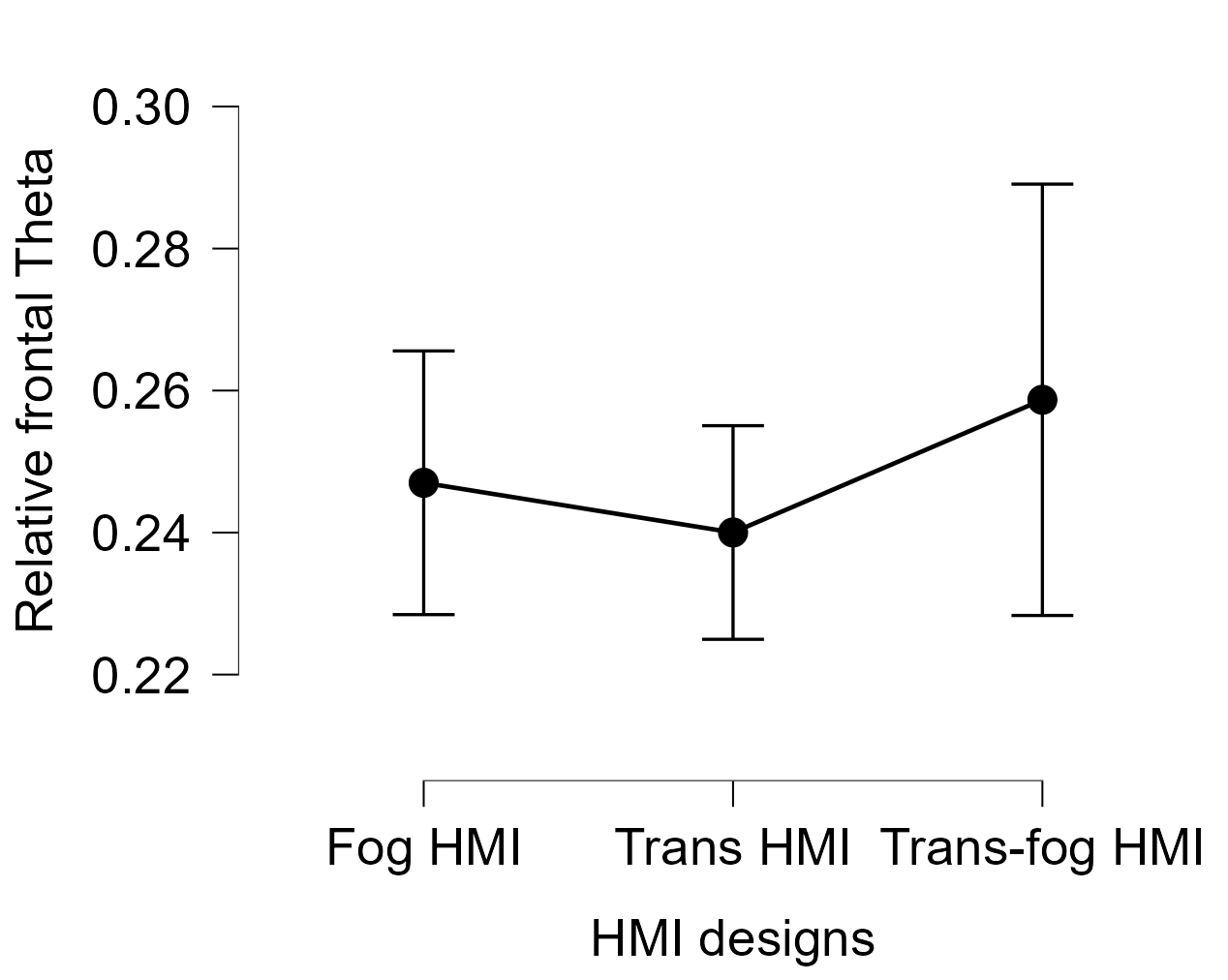}
         \caption{Relative Theta power.}
         \label{fig:theta_power}
     \end{subfigure}
        \caption{EEG power spectral analysis with mean values and standard errors of means.}
        \label{fig:EEG_power}
\end{figure}

In Fig.~\ref{fig:EEG_power} we can see the power spectral analyses of relative Alpha band power and Theta band power. We can see from Fig.~\ref{fig:EEG_power} that the Trans HMI, which we expected to have the lowest mental workload, had the highest mean relative Alpha power. Although this result was in agreement with the literature, where decreased mental workload is associated in increased Alpha power, but the difference was not significant $F(2,18)=0.55, p=0.58, \eta_{p}^{2}=0.058$. Similarly for relative Theta band power, despite the Trans HMI owned the lowest mean value, implying minimum effort was required to understand it, no significant difference was found $F(2,18)=0.18, p=0.83, \eta_{p}^{2}=0.02$. 

\begin{table}[htbp]
\caption{Descriptive Data for All Workload Measures and Subjective Transparency.}
\centering
\begin{tabular}{ccccc}
\toprule
\multicolumn{1}{c}{} & \multicolumn{2}{c}{\textbf{EEG power}} & \multicolumn{2}{c}{\textbf{Subjective}} \\
\multicolumn{1}{c}{} & \multicolumn{2}{c}{\textbf{spectral}} & \multicolumn{2}{c}{\textbf{measurements}} \\
\cmidrule(rl){2-3} \cmidrule(rl){4-5}
\textbf{\makecell{HMI\\designs}} & \makecell{relative\\Alpha} & \makecell{relative\\Theta} & {NASA-TLX} & \makecell{Subjective\\transparency} \\
\midrule
Fog & 0.186 (0.082) & 0.247 (0.053) & 35.4 (21.2) & 55.5 (18.5)  \\
Trans & 0.195 (0.107) & 0.240 (0.059) & 28.9 (17.7) & 79.8 (16.2)  \\
Trans-fog & 0.183 (0.074) & 0.259 (0.146) & 37.1 (19.7) & 74.9 (18.3)  \\
\bottomrule
\end{tabular}
\label{table:descriptive}
\end{table}

\section{Discussion}\label{discus}
In this study, we were eager to determine whether different mental workloads could be found in those three developed HMI designs and if the EEG spectral power analysis could be used to find the differences. This should be deemed as a starting point in developing a systematic and standardized HMI design assessment method, where we intend to incorporate psychophysiological measures into the proposed transparency assessment test, to make the HMI assessment process more objective and efficient.

Three HMI designs were developed and validated based on heuristics and results from the previous study, to have differences in the understandability and the easiness to understand them. The subjective mental workload measured using NASA-TLX confirmed Hypothesis 1. Significantly lower workloads were required by participants to understand the information transmitted by the L2 automated driving system (ADS) when using the Trans HMI. This HMI design was granted with both HMI and system transparency, which means that on the interface side, the information shown on the Trans HMI could be understood with minimum effort; while on the system side, the logic behind L2 automation activation on the Trans HMI is clearer: L2 automation could only be activated when it is available. 

During the feedback section, a common reason for L2 ADS owners to turn the L2 function off was that they had no idea if it was on or off. Since users remain responsible throughout the whole time during L2 AV driving, most of the HMI designs expect users to monitor whether L2 is activated the whole time. Hence, even if the activation of L2 automation failed, instead of returning to the previous state and warning the users, the ADS is designed to enter the standby mode, where the lateral control will be automatically activated whenever the system is ready. And we consider this type of system logic intransparent, or as we put in the name of Fog HMI, foggy. The effect of the system transparency on mental workload was confirmed by the pairwise comparison between the Trans and Trans-fog HMIs, where it was the only difference between the two.

The results from the subjective transparency test confirmed Hypothesis 2, and it acted in accordance with the results from NASA-TLX, where Trans HMI remained to be the most transparent, i.e., easiest to understand, HMI design, given that it had the highest mean subjective transparency score. What we can note here is that the Fog and Trans-fog HMIs had around the same NASA-TLX score, but when it comes to subjective transparency score, the Trans-fog HMI had significantly higher values than that of Fog HMI. This could be explained by this additional construct included in the concept of subjective transparency, which is the capability of obtaining the information needed. When evaluating the subjective transparency, besides being asked if they agreed that the HMI design is easy to understand (which is a similar construct as workload), participants were asked if they agreed that they could obtain critical information from the HMI design. By design, the Trans-fog HMI had the advantage of HMI transparency, which allowed users to obtain correct information and thus resulted in higher subjective transparency.

After the confirmation of Hypothesis 1 and 2, we identified three HMI designs with differences in mental workload. Since workload plays an important role in estimating the transparency of HMI designs \cite{Liu2022}, it is urgently required for us to identify an objective workload measurement to make the proposed method efficient in real driving scenarios. Due to the high temporal resolution and ability to measure neural activities directly, EEG was chosen to be the psychophysiological measure in this study. However, Hypothesis 3 was not confirmed by the results. The possible reason is that the sample size was too small. Due to the loss of recording, the number of usable data dropped, which might explain why the statistical power is lower than expected. A possible solution besides recruiting more participants is to not combine all the epochs as one for each HMI design but keep them separated. When analyzing the spectral power of EEG, data with longer duration could average out the noises, which is why we combined all the epochs around the activations of L2 into one. Keeping them separated could, on the other hand, increase the sample size, and in return, increase the statistical power.

It might also be possible that it is difficult to identify differences in mental workload among HMI designs using the analysis of the spectral power of EEG. In this case, we should consider other psychophysiological measures to extend the transparency assessment method into real driving scenarios. But before that, more studies have to be done to come to that conclusion.

\section{Conclusions and Future Work}\label{conclud}
In conclusion, the first part of the research question was successfully answered and confirmed by the results. This study proposed and validated three HMI designs with differences in mental workload and subjective transparency. We can also learn from the results that, in order to make the HMI design more understandable and easy to use, being only transparent on the interface might not be sufficient, as system transparency also plays a big part in it. More works and discussions on the logic behind the HMI and the ADS are necessary.

This study can also be regarded as the basis for the development of the transparency assessment method. In the future study, we would utilize the validated HMI designs to identify the psychophysiological measures that are effective in identifying differences in mental workload and transparency in real driving scenarios. By doing so, we could make the HMI more transparent, and the riding in automation safer.

\newpage

\bibliographystyle{IEEEtran}
\bibliography{IEEE}

\end{document}